**Emergence of mixed orientational ordering in quasi-one-dimensional superdisk and superball fluids**

*Sakineh Mizani[a), Martin Oettel[a)], Péter Gurin[b)] and Szabolcs Varga*[b)]*

[a)]*Institute for Applied Physics, University of Tübingen, Auf der Morgenstelle 10, 72076 Tübingen, Germany*

[b)]*Physics Department, Centre for Natural Sciences, University of Pannonia, PO Box 158, Veszprém, H-8201 Hungary*

*corresponding author:vargasz2025@gmail.com

**Abstract**

We report the discovery of a mixed orientational structure in the quasi-one-dimensional fluid of hard non-spherical bodies with the exact calculation of the thermodynamic and structural quantities using the transfer operator method. The mixed arrangement, which is spatially uniform, but orientionally ordered, cannot be identified with conventional mesophases such as tetratic, cubatic and nematic. It is found that the particles form a mixed orientational arrangement with preferred parallel and perpendicular orientations in a channel, where the number of parallel and perpendicularly oriented particles is not equal even at the close packing density. The mixed structure can be stabilized with hard bodies having equal side lengths in parallel and perpendicular orientations along the channel. These conditions can be realized with colloidal superdisks (superballs) if the curvature of neighboring sides (faces) are different. We show that even a small stretching of the superparticle destabilizes mixed ordering due to perfect nematic order evolving upon approaching close-packing.

**Introduction**

Bottom-up assembly of complex functional materials requires colloidal building elements whose forms, surface characteristics, and directional interactions can be accurately controlled [1-3]. Over the last two decades, advances in particle synthesis have widened the avenue for building blocks, providing unprecedented control over their anisotropy in shape, surface chemistry, and bonding directionality [4-8]. Colloidal particles with specific DNA-coating form patchy colloids and can now be programmed to assemble into very complex structures including solid structures such as cubic diamond lattices and Laves phases [9]. It is also possible to make colloidal particles anisotropic in shape using for example gradient-stretching technique [10]. Polymeric colloids with a wide range of aspect ratios can be realized [11, 12], while novel surface-engineering methods enable to add chemical patchiness to anisotropic colloidal particles [13]. Moreover, hard colloidal particles with engineered shapes ranging from convex forms like the polyhedral and plate-like particles to concave shapes like the star and lens particles are also available. In these systems the entropic forces play the key role in the ordering and self-assemble [14-27]. The advances made in the shape engineering are positioning the anisotropic colloids as the "atoms" and "molecules" of tomorrow's materials, opening new routes to assemble photonic crystals, metamaterials, and other complex architectures [28-31].

In many studies of colloidal self-assembly, colloidal particles are idealized as perfect sphere, cube, or ellipsoid in shape. In reality, colloids frequently deviate from these idealized shapes, and even subtle modifications such as corner rounding, edge curvature, or slight changes in aspect ratio can profoundly alter phase stability and transformation pathways. In line with simulation and experimental studies, these non-ideal features can sometimes lead to anomalous ordering and phase transitions, as well as the formation of phases that are not present in "ideal" limits [32-40]. These shape-dependent effects are evident across different dimensions. For example ideal hard cubes in three dimensions do not have rotator phases, while hard superballs, which are between cubes and spheres have stable rotor phases with face-centered cubic and body-centered cubic structures throughout a wide range of shape [41]. Beyond ordered phases, the structure of disordered packings is likewise shape-sensitive. For example, the density of maximally random jammed superball packings changes dramatically and non-analytically with moving away from the spherical limit, which reflects fundamental differences in local packing arrangements between spherical and non-spherical particle shapes [34]. Similar amazing shape effects occur in two-dimensional (2D) systems. Ideal 2D hard squares form tetratic and square crystal structure upon compression. However, experiments on monolayers of colloidal hard platelets show that the particles actually form hexagonal rotator and rhombic crystals at high densities [42]. Varying the corner curvature of hard squares from circular to perfectly square reveals the existence of different ordered phases. Depending on roundness, the isotropic fluid changes to hexagonal rotator phase (via intermediate hexatic) or square phase (via intermediate tetratic). Moreover, in a narrow range of corner curvature, a polycrystalline state exists, where square domains and weakly hexagonal clusters coexist [43]. A generalized class of rounded squares, known as superdisks, can display even richer ordering behavior than conventional shapes, exhibiting an unusual three-step melting scenario—rather than the typical two-step process—mediated by both hexatic and rhombatic phases, as revealed by replica-exchange Monte Carlo simulations [44]. The shape sensitivity extends to three-dimensional (3D) anisotropic hard body systems, too. For example, hard cylinders can form smectic-A phases upon compression [45] that are strictly absent in hard ellipsoid systems [46]. This fact further demonstrates the fundamental role played by the geometric details of the particle's shape in the self-assemble and phase transitions. While these effects have been well-documented in 2D and 3D systems, their manifestation in quasi-one-dimensional (q1D) systems remains largely unexplored.

Due to its simplicity, q1D confinements have emerged as a powerful platform to probe fundamental aspects of structure and dynamics in colloidal fluids [47-52]. In highly confined q1D environments such as cylindrical channels or slit-like pores, the confinement can influence both the positional and orientational ordering of colloids substantially [53-56]. When colloids are confined within channels whose widths approach the particle size, spatial restrictions and excluded volume interactions together produce dramatic deviations from 3D bulk behavior [57]. Experimental studies on hard-sphere colloids in ribbon-like microchannels have revealed phenomena such as transverse stratification, anisotropic pair correlations, and liquid-like ordering along layers that resist full crystallization even at high densities [58]. Specifically, the pair correlation functions in different strata exhibit anisotropic behavior, while the longitudinal order along the strata remains characteristic of a liquid rather than a crystalline phase [59]. Particle motion in these q1D systems exhibits a crossover from short-time normal diffusion to long-time single-file diffusion, where geometric confinement prevents mutual passage of particles [60, 61]. The resulting subdiffusive scaling of mean-square displacement has been confirmed through optical tweezer experiments [62] and described theoretically with short-time fluctuation analyses [63]. Simulations of hard disks confined in narrow channels demonstrate emergent zig-zag ordering, optical-like transverse modes, and specific dispersion laws for collective excitations that are intermediate between 1D and 2D systems [64, 65]. Further exploration in circular channels reveals cascades of diffusion regimes controlled by particle interactions and external fields, illustrating the rich interplay between curvature, topology, and confinement [66].

As the experimental techniques do not produce colloids with ideal geometrical shapes such as spheres and ellipsoids, the deviation from the perfect shape can be described with generalized geometrical models having some extra parameters, which makes possible to tune the shape between ideal geometrical objects [33, 67]. One example is the superdisk model, which interpolates between squares and disks with a deformation parameter tuning the roundness of the disk's side [68]. This model can be useful for plate-like colloidal particles confined into a flat surface, where the particle's bottom and top surfaces are parallel with the confining surface and they collide with their edges in the 2D plane. The 3D generalization of this model is the superball, which allows us to interpolate between sphere and cube with the help of a deformation parameter [69]. If the colloidal particle is elongated, it can be modeled as an ellipsoid, which can be made both plate-like and rod-like with varying its aspect ratio. The generalization of the superball model for non-spherical colloidal particles is the superellipsoid model depending on the aspect ratio and deformation parameter. With this superellipsoid model it is possible to study colloids, which can be plate-like, almost spherical and rod-like in shape with the extra property that the curvature of the faces can be varied through the deformation parameter [70]. To include the special feature of all faces and sides of the colloidal particles, it is possible to extend the above models with the generalization of the superellipse and superellipsoid models with using more than one deformation parameter. Using two deformation parameters, the curvature of different sides and faces of the colloidal particles can be tuned independently, i.e. one side can be more curved than the other. In our study we examine how the side and surface curvatures affect the orientational ordering of colloidal hard bodies confined into narrow channels.

Although 2D and 3D experiments show that very weak non-sphericity of the particle's shape affects mainly the crystalline ordering of colloidal particles, q1D confinements can magnify the importance of the geometrical details of the particle's shape both in dilute and dense systems [71]. Here we show that the tiny deviation from the spherical shape can result in an unusual ordering of almost spherical colloidal hard particles if the particles are confined into a very narrow hard channel. If the curvature of the particle deviates a bit from the sphere's constant curvature in some directions, a mixed ordering can be stabilized, which cannot be categorized as nematic, cubatic or tetratic ordering as two preferred orientations are present in the whole density range, where one of two orientations is more preferred than the other. We use the generalized superdisk and superball models to prove the existence of the mixed structure. Note that the hard superdisk model with two deformation parameters is suitable to study ordering of colloidal plates moving on a very narrow flat stripe. With this model, we study the effect of curvature changes in the cross section of the colloid plate on the orientational ordering. The generalized hard superball model is used to study the ordering of non-ideal colloidal spheres confined into a cylindrical nanochannel. With varying the curvature of neighboring faces independently, we show that the orientational ordering of the particle's faces is not isotropic along the cylindrical channel in the entire density range. Our results show that both colloidal platelets and balls can form mixed ordering. Furthermore, we study the stability of the mixed ordering with respect to changing the aspect ratio of the superparticle. Our results show that the mixed structure is very sensitive to the aspect ratio of the particle.

**Quasi-one-dimensional superparticle models**

In order to show that tiny deviation from the spherical shape results in macroscopic effects in the ordering properties of quasi-one-dimensional (q1D) colloidal systems, we consider two hard superparticle models. The first model is the two-dimensional (2D) superdisk, while the second one is the three-dimensional (3D) hard superball, which are defined by

$$\left|\frac{x}{a}\right|^n + \left|\frac{y}{b}\right|^m \leq 1, \tag{1}$$

and

$$\left|\frac{x}{a}\right|^n + \left|\frac{y}{a}\right|^n + \left|\frac{z}{b}\right|^m \leq 1. \tag{2}$$

Here $a$ and $b$ are the semi-axis lengths along the Cartesian axes, while $n$ and $m$ are deformation parameters. Regarding the superdisk shape, Fig. 1 shows that the role of $n$ and $m$ is to control the roundness of the sides of the particle. If $n=m$, then all sides have the same roundness. Especially, $n=m=2$ corresponds to disk, while $n=m\to\infty$ corresponds to square. The case of $n\neq m$ gives rise to different roundness for the neighboring sides, while the roundness of opposite sides remain the same. It can be seen in Fig. 1 that the four-fold rotational symmetry of the superdisk shape changes to be two-fold one if $n \neq m$. With choosing $d=2a$ to be the unit of the length, the shape of the particle can be changed to be a rod-like superellipse via the aspect ratio, which is defined as $k=b/a$.

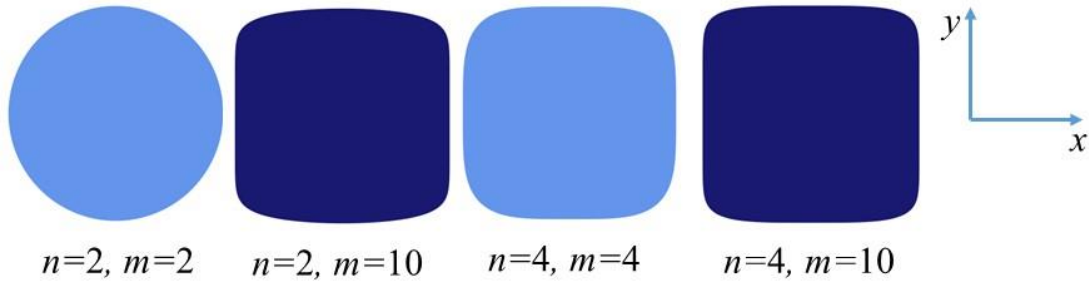

$n=2, m=2$   $n=2, m=10$   $n=4, m=4$   $n=4, m=10$

**Figure 1.** Effect of the deformation parameters $n$ and $m$ on the shape of the hard superdisk particle ($k=1$) according to Eq. (1). The particle corresponds to the hard disk for $n=m=2$. The roundness of left and right sides can be varied with $m$, while $n$ is responsible for the roundness of upper and lower sides.

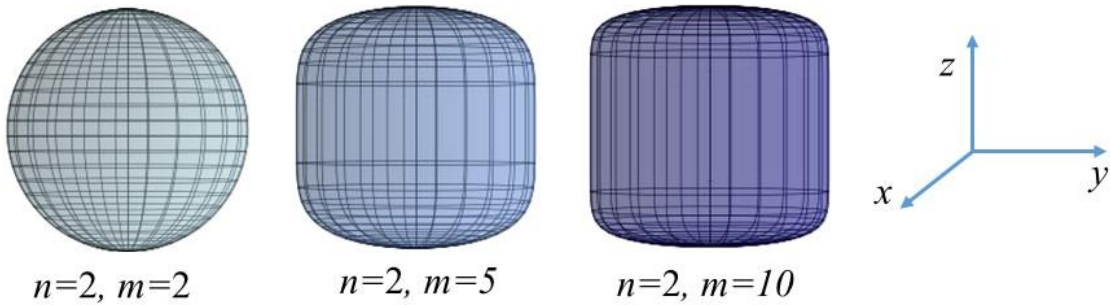

$n=2, m=2$   $n=2, m=5$   $n=2, m=10$

**Figure 2.** Effect of the deformation parameters $n$ and $m$ on the shape of the hard superball ($n=2$ and $k=1$) according to Eq. (2). The particle corresponds to the hard sphere if $m=2$. The roundness of the side surface along $z$ axis can be varied with $m$, while the roundness of top and bottom surfaces is always quadratic as $n=2$. Note that the particle is uniaxial in shape because of the continuous rotational symmetry around the $z$ axis.

Therefore, $k=1$ corresponds to a superdisk, while $k>1$ to a superellipse. In this study we examine the effect of tiny deviation of $k$ from 1 on the ordering properties of q1D superdisks if $n \neq m$. In the case of superballs, we prescribe the uniaxial symmetry around $z$ axis in the shape using the same exponent $n=2$ for $x$ and $y$ variables, while $k=1$ and $m \geq 2$. With these choices, the shape of superball can be varied between spheres and rounded cylinders, which is shown in Fig. 2. We can see that the roundness of the upper and lower surfaces of the superball is always quadratic as $n=2$, while $n=2$ and $m$ together are responsible for the roundness of the side surface of the superball. As $k=1$, the length of the superball is the same along all Cartesian axis. In this study we examine the effect of varying $m$ on the orientational ordering properties of q1D hard superballs.

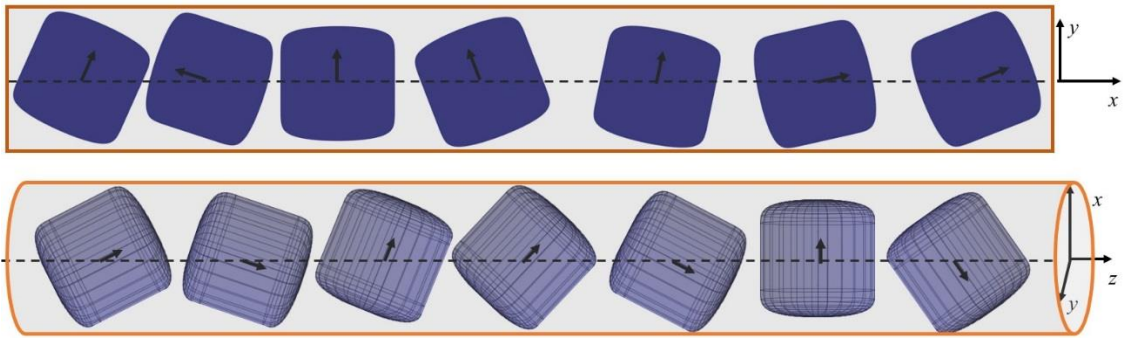

**Figure 3.** Quasi-one-dimensional system of hard superdisks and that of hard superballs in the upper and lower panels, respectively. The superdisks are on a stripe, while the superballs are in a cylindrical channel. The arrows represent the orientation of the particles. The superdisks are rotating freely in the 2D $x$-$y$ plane, while the superballs in the 3D x-y-z space. The out-of-line positional fluctuations are not taken into account, i.e. the centres of the particles are allowed to move freely only on the central dashed line. The particles are characterized with $k=1$, $n=2$ and $m=10$.

In this study we assume that only excluded volume interactions act between the colloidal particles, i.e. overlaps between superparticles are not allowed. Moreover, superparticles are placed into a q1D environment. This means that hard superdisks are constrained to move in a stripe, while hard superballs are placed into a cylindrical channel, where the particles are not allowed to pass each other and only first neighbor interactions are present. As the stripe and the channel are very narrow, we do not take into account the effect of transversal positional fluctuations. However, the particles are allowed to freely move and rotate along the channel. In the hard superdisk fluid, the particles can rotate in two dimensions with the angle $\varphi$ measured from the $y$ axis. Note that it is enough to consider only $-\pi/2 < \varphi < \pi/2$ interval due to the apolar shape of the superdisk body. The orientation of a uniaxial superball is characterized with a polar ($\theta$) and an azimuthal angle ($\varphi$). As the particle has up-down symmetry we consider only the following orientations: $0 < \theta < \pi/2$ and $0 < \varphi < 2\pi$. One possible realization of q1D superdisk and superball systems are shown in Fig. 3.

**Exact method for bulk and orientational properties**

The statistical mechanical treatment of many systems can be done exactly using the well-known transfer operator method if only few (first, second, etc.) neighbor pair interactions are present [72]. One class of exactly solvable models is the fluid of hard bodies confined into narrow channel [73]. As our superparticle systems belong to this class, we use this method in *NPT* ensemble, where $N$ is the number of superparticles, $P$ is the longitudinal pressure and $T$ is the temperature. The basic idea of the method is to rewrite the configuration part of the isobaric partition function ($Z$) as a trace of matrix products such that $Z = TrK^N$, where the matrix product is defined as $K^2(I_{i-1}, I_{i+1}) = \int dI_i K(I_{i-1}, I_i) K(I_i, I_{i+1})$. In this matrix product $K(I_i, I_{i+1})$ is the kernel function coming from hard body exclusion between particles $i$ and $i+1$, where $I_i$ and $I_{i+1}$ are the set of configuration variables of the neighboring colloidal particles. As the result of trace operation is independent of the basis used, it is useful to work in the eigenfunction basis of the kernel-function, which is defined by the following eigenvalue-equation

$$\int dJ\, K(I,J)\psi_n(J) = \lambda_n \psi_n(I), \tag{3}$$

where $\lambda_n$ is one of the eigenvalues and $\psi_n$ is the corresponding eigenfunction. To solve Eq. (3), it is prescribed that the eigenfunctions are normalized and orthogonal, i.e.

$$\int dJ\, \psi_i(J)\psi_j(J) = \delta_{ij} \tag{4}$$

where $\delta_{ij}$ is the Kronecker delta function. The largest eigenvalue ($\lambda_0$) and the corresponding eigenfunction ($\psi_0$) determine the bulk and orientational properties. The Gibbs free energy ($G$), the equation of state (density-pressure relationship) and the orientational distribution function ($f$) are given by

$$\beta G/N = -\ln \lambda_0, \tag{5a}$$

$$1/\rho = \frac{-1}{\lambda_0}\frac{d\lambda_0}{d\beta P}, \tag{5b}$$

and

$$f(I) = \psi_0^2(I). \tag{5c}$$

Here $\beta=1/k_B T$ is the inverse temperature, and $\rho=N/L$ is the one dimensional density, where $L$ is the length of the channel. As the pair interaction is hard exclusion, i.e. particles are not allowed to overlap, the kernel function of the superparticles depends on the contact distance between two neighboring particles having $I$ and $J$ coordinates ($\sigma(I,J)$). It can be shown by integrating out the longitudinal positional variables of the particles that

$$K(I,J) = \exp(-\beta P \sigma(I,J))/\beta P. \tag{6}$$

Using $\sigma(I,J)$, which can be obtained using simple geometry, the eigenvalue-equation (Eq. (3)) together with Eqs. (4) and (6) can be solved with numerical integration and successive iteration for a given value of $P$. Using the largest and the second largest eigenvalues ($\lambda_0$ and $\lambda_1$), the orientational correlation length can be obtained from

$$\xi^{-1} = \ln(\lambda_0/\lambda_1), \tag{7}$$

which is measured as a function of the number of particle between two particles and not of distance [49].

Regarding $I$ and $J$, they correspond to the angles of particle's orientations. While the orientation of a superdisk is defined by an angle $\varphi$, the orientation of a superball is defined by angles $\varphi$ and $\theta$. The contact distance between two superdisks depends on the angles of two particles, $\sigma=\sigma(\varphi_1,\varphi_2)$. Therefore, the eigenfunctions of superdisk fluid are angle dependent as follows $\psi_n = \psi_n(\varphi)$. In the case of superballs, the angle dependence of contact distance and the eigenfunctions can be written as $\sigma=\sigma(\theta_1,\theta_2,\varphi_{12})$ and $\psi_n = \psi_n(\theta,\varphi)$, where $\varphi_{12}=\varphi_1-\varphi_2$ is the relative azimuthal angle. Note that actually $\psi_0 = \psi_0(\theta)$, i.e. there is no in-plane order in the x-y plane [53]. This is consistent with the expectation that the orientational ordering of uniaxial superballs must be uniaxial along the channel.

As all particles are parallel with their shortest length along the channel at the close packing density, the fluid of superparticles without orientational freedom is an interesting reference system to understand the close packing properties of freely rotating particles. Luckily, the system of parallel superparticles without orientational freedom can be treated analytically, because in this case $K = \exp(-\beta P d)/\beta P$ is a simple number (one by one matrix), where $d=2a$ is the shortest distance between two parallel superparticles. In this parallel system the largest eigenvalue of Eq. (3) is proportional to the kernel function, i.e. $\lambda_0 \sim \exp(-\beta P d)/\beta P$. Using Eq. (5b) we get that $1/\rho = d + 1/\beta P$, which can be re-expressed as $\beta P = \rho/(1-\rho d)$. From this equation we can obtain the close packing density ($\rho_{cp}$) in $\beta P \to \infty$ limit, which is given by $\rho_{cp} = 1/d$. In the case, when all particles are parallel, the pressure is denoted as $P_\parallel$, i.e.

$$\beta P_\parallel = \rho/(1-\rho d). \tag{8}$$

To see the effect of orientational fluctuations on bulk properties in the vicinity of the close packing density, we determine $P/P_\parallel$, where $\rho$ coming from Eqs. (3), (4), (5b) and (6) is used to obtain $P_\parallel$. The normalization condition for the orientation distribution function differs for superdisks and superballs because of the difference in their orientation freedom. Starting from Eqs. (4) and (5c) the normalization of $f$ for superdisks and superballs can be written as

$$\int_{-\pi/2}^{\pi/2} d\varphi\, f(\varphi) = 1, \tag{9a}$$

$$\int_0^{\pi/2} d\theta \sin\theta \int_0^{2\pi} d\varphi\, f(\theta) = 1. \tag{9b}$$

As $f$ does not depend on $\varphi$ in the fluid of superballs, i.e. $f=f(\theta)$, the integration with $\varphi$ gives a prefactor of $2\pi$ in Eq. (9b). Based on the normalization condition of $f$, we can determine the fraction of superdisk particles with vertical orientation. We consider the particles to be vertical, if they are within $-\pi/4 \leq \varphi \leq \pi/4$, while the particles are horizontal with angles $\pi/4 < |\varphi| \leq \pi/2$, we can write the fraction of vertical particles as

$$x_v = \int_{-\pi/4}^{\pi/4} d\varphi\, f(\varphi). \tag{10}$$

Similarly, the fraction of radially oriented superball particles can be obtained from

$$x_r = 2\pi \int_{\pi/4}^{\pi/2} d\theta \sin\theta\, f(\theta), \tag{11}$$

where particles within $\pi/4 \leq \theta \leq \pi/2$ are considered to be radial, while particles in $0 \leq \theta \leq \pi/4$ are axial. With this definition it is guaranteed that the sum of vertical (radial) and horizontal (axial) particles is always $N$ in accordance with the normalization condition of the orientational distribution function.

**Results**

Now we present the bulk and orientational properties of q1D superdisk fluids, which are obtained numerically using the transfer operator method. As the particles are not allowed to overlap, i.e. only excluded volume interactions are present, the distance between two particles cannot be lower than the contact distance. Actually, the contact distance (appearing in the kernel function, see Eq (6)) is the crucial element for the method. Therefore, we show the orientation dependence of the contact distance in Fig. 4 for superdisks with deformations $n=2$ and $m=10$. Since $k=1$, i.e. $a=b$, the contact distance is minimal with $\sigma=2a$ for vertical-vertical ($\varphi_1=\varphi_2=0$), horizontal-

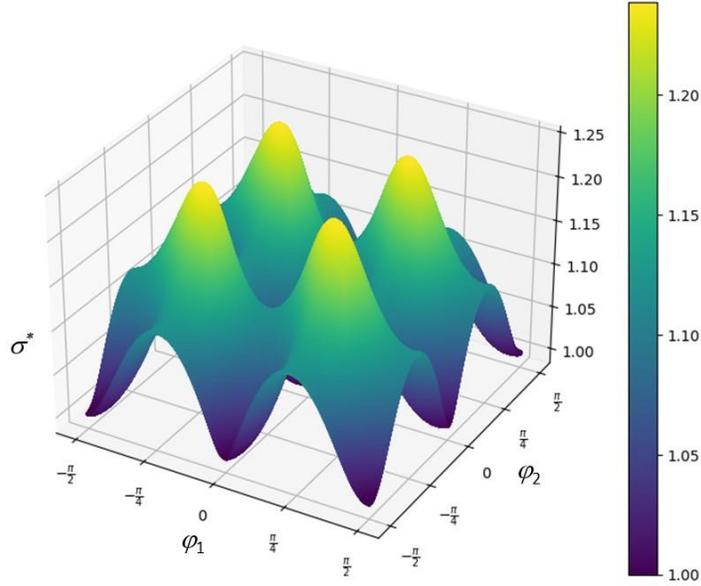

**Figure 4.** Contact distance between two superdisks having $\varphi_1$ and $\varphi_2$ angles. The shape of the superdisk is characterised with $n=2$ and $m=10$. $\sigma^*=\sigma/d$ is the dimensionless contact distance.

-horizontal ($\varphi_1=\varphi_2=\pm\pi/2$) and vertical-horizontal ($\varphi_1=0$ and $\varphi_2=\pm\pi/2$ or $\varphi_1=\pm\pi/2$ and $\varphi_2=0$) orientation pairs of two superdisks. Therefore nine orientation pairs give the same minimum in the contact distance vs. $\varphi_1$-$\varphi_2$ surface, which are separated by peaks in Fig. 4. In the light of contact distance surface it is very hard to predict the orientational ordering of the superdisks upon compression, because both horizontal and vertical orientations can be equally good to maximize the available room for the particles. In other words, the close packing structure is degenerate, i.e. it can be realized with many configurations. Therefore, those configurations produce the equilibrium structure, which have the lowest contact distance cost during angular fluctuations. The consequence of separated minima in the $\sigma(\varphi_1,\varphi_2)$ surface is that the system can be trapped easily into non-equilibrium states such as glassy and jammed states upon sudden compression [50, 51, 74].

In the following figures we show that the equilibrium structure of q1D superdisks is neither nematic nor tetratic, but it is a mixed structure. To understand the unusual behavior of q1D superdisk fluids, we show the main features of nematic, tetratic and mixed structures in Fig. 5, where the effects of deformation parameters (*n* and *m*) and aspect ratio (*k*) on the orientational distribution function are presented. We can see that the ordering is tetratic with orientational ordering around the angles $\varphi=0$ and $\varphi=\pm\pi/2$ for $k=1$ and $n=m=4$. Moreover, $f(0)=f(\pm\pi/2)$ is due to the fourfold symmetry of the superdisk shape. The consequence of tetratic ordering is that the number of vertical and horizontal particles is always equal, i.e. $x_v=x_h=1/2$. The tetratic ordering can be characterized with the order parameter $S_4=|<\cos 4\varphi>|$ alone, because $S_2 =|<\cos 2\varphi>|$ is always 0. It can be shown that $S_4=0$ for the isotropic distribution ($f=1/\pi$) and $S_4=1$ for the perfectly ordered tetratic one. The fourfold symmetry of the particle's shape changes to be twofold for rod-like shapes ($k>1$). This case is shown for superellipses with $k=3$ and $n=m=4$ in Fig. 5. In this case, the ordering is nematic, because *f* is single-peaked at $\varphi=0$ with the consequence that $S_2>0$ and $S_2$ goes 1 in the close packing (high pressure) limit. The four-fold rotational symmetry can be also broken into two-fold one with $n\neq m$ even if $k=1$. The resulting ODF is shown for superdisks with $n=2$ and $m=10$ in the lower row of Fig. 5. We can see that $f(0)\neq f(\pm\pi/2)$ and both $f(0)$ and $f(\pm\pi/2)$ do not vanish at very high pressures. The consequence of these results is that $S_4$ goes to 1 in the close packing limit, but $S_2$ not. The other import feature of this arrangement is that $x_v$ is always larger than 0 and less than 1/2. Therefore, this arrangement cannot be classified as either tetratic or nematic, i.e. it is of mixed order. In Table 1 the main features of tetratic, nematic and mixed ordering are summarized.

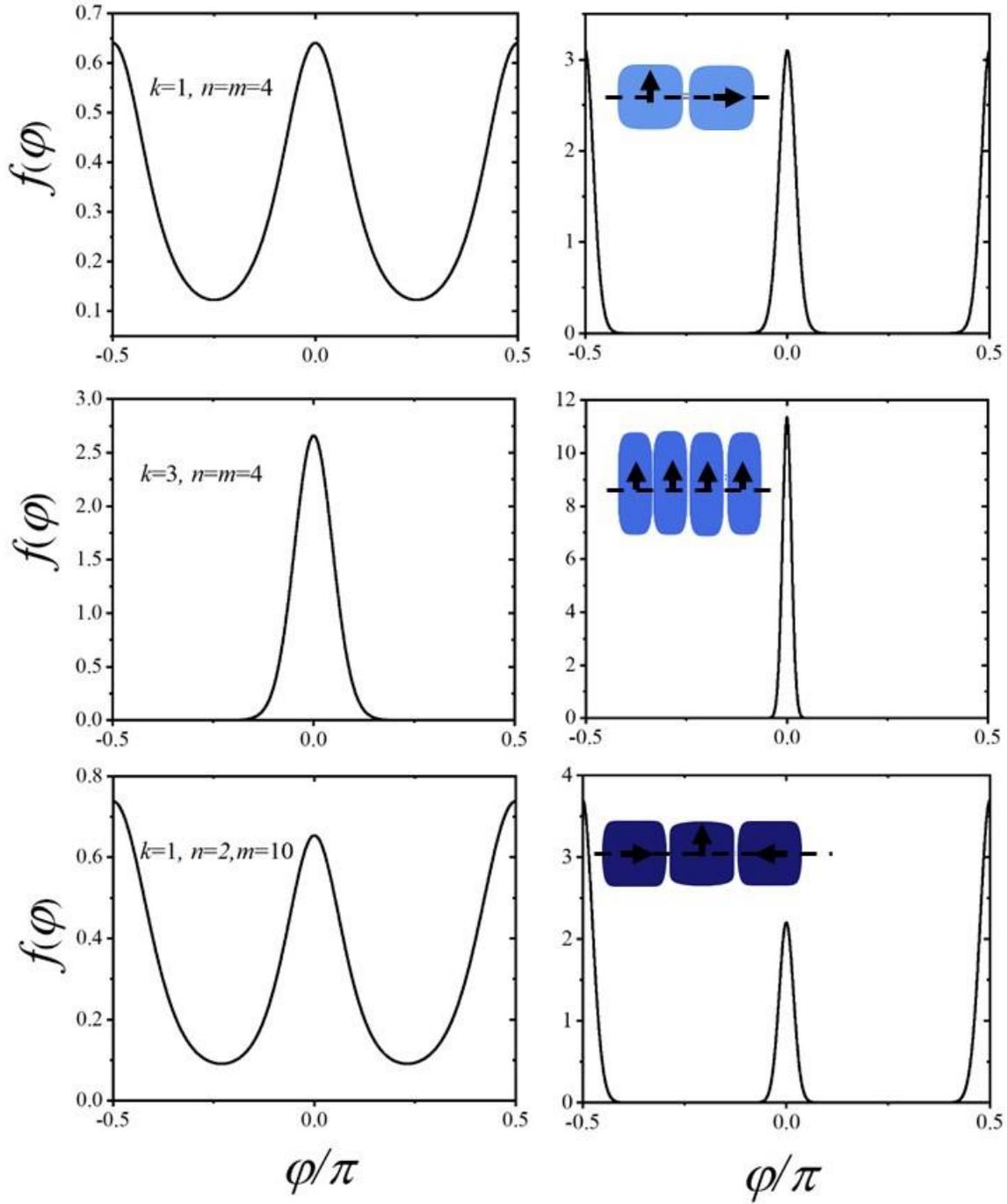

**Figure 5**. Orientational distribution function of tetratic (upper panels), nematic (middle panels), and mixed (lower panels) phases at pressures $P^*=10$ (left column) and $P^*=100$ (right column). The corresponding close-packing (high-pressure) structures are also shown. $P^*=\beta Pd$ is the dimensionless pressure.

**Table 1.** Order parameters ($S_2$ and $S_4$), fraction of vertical particles ($x_v$) and the ratio of orientational distribution function at vertical and horizontal orientations ($f(0)/f(\pi/2)$) of the nematic, tetratic and mixed orientational ordering at intermediate and close packing (*cp*) densities

| tetratic | $S_2=0$ | $0<S_4<1$ | $x_v=1/2$ | $S_{2,cp}=0$ | $S_{4,cp}=1$ | $x_{v,cp}=1/2$ | $f(0)_{cp}/f(\pi/2)_{cp}=1$ |
|---|---|---|---|---|---|---|---|
| nematic | $0<S_2<1$ | $0<S_4<1$ | $1/2<x_v<1$ | $S_{2,cp}=1$ | $S_{4,cp}=1$ | $x_{v,cp}=1$ | $f(0)_{cp}/f(\pi/2)_{cp}=\infty$ |
| mixed | $0<S_2<1$ | $0<S_4<1$ | $1/2<x_v<1$ | $S_{2,cp}<1$ | $S_{4,cp}=1$ | $x_{v,cp}<0.5$ | $f(0)_{cp}/f(\pi/2)_{cp}<1$ |

The existence of mixed order cannot be understood with the help of Fig. 4 due to the presence of 9 minima in the contact distance. Therefore we have to study more details of the contact distance at some angles. In Fig. 6 we show the contact distance at $\varphi_1=0$ and $\varphi_1=\pi/2$ if the orientation of the neighboring second particle is changing. For $k=1$, the increment of the contact distance from the minimal contact value is quadratic in $\varphi_2$ around the angles $\varphi_2=0$ and $\varphi_2=\pm\pi/2$ if the particle 1 is horizontal ($\varphi_1=\pm\pi/2$). This shows that a horizontal particle likes to be the neighbor of both horizontal and vertical particles. If the particle 1 is vertical ($\varphi_1=0$), the contact distance increases quadratically with $\varphi_2$ around $\varphi_2=\pm\pi/2$, but this dependence changes to be linear in $\varphi_2$ around $\varphi_2=0$. This means that a small angular fluctuation has lower contact distance "cost" in the case of the vertical-horizontal neighborhood than in the case of the vertical-vertical one (see the curves of $k=1$ cases in the vicinity of $\varphi_2=0$ in Fig. 6). At high pressure only small angular fluctuations occur, therefore the vertical particles likes to be the neighbor of horizontal but not vertical ones, while the horizontal particles equally like horizontal and vertical neighboring particles. To minimize the cost of angular fluctuations in the contact distance and to maximize available room for all particles, the number of horizontal particles exceeds the number of vertical ones upon compression. Moreover, vertical particles are surrounded mainly by horizontal ones, while horizontal particles do not select between vertical and horizontal ones to be their neighbors. This indicates that vertical particles have to be present in very dense systems, surrounded by horizontal ones, i.e. the cluster of vertical particles cannot evolve. With the presence of vertical particles in the entire density range, the number of different orientational states can be increased, which definitely increases the orientational entropy, but does not affect the packing entropy. As the result of the entropic game, the mixed ordering is the stabilized instead of tetratic and nematic structures. The case of $k>1$ is different, because the degeneracy of the contact distance in horizontal and vertical orientations is broken with favoring the vertical orientations. Fig. 6 shows that the contact distance has a minimum for vertical-vertical neighboring pairs, while both the vertical-horizontal and the horizontal-horizontal neighbors have higher contact distance. As the driving force of the ordering is to minimize the occupied length along the channel, the rod-like particles must form nematic ordering at high densities.

In the followings we examine the competition between mixed and nematic ordering upon varying the aspect ratio of the superparticle. The pressure ratio between freely rotating and parallel superballs is shown as a function of density in Fig. 7. In the case of superdisks with $n=2$ and $m=10$, $P/P_\parallel$ is single-peaked at an intermediate density and converges to $\alpha=1.5$ at $\rho^*=1$, where $\alpha$ is the limiting value of $P/P_\parallel$. The peak indicates the continuous change from isotropic to mixed ordering, while $\alpha=1.5$ shows that the angular fluctuations still have macroscopic contribution with $P_\parallel/2$ to the pressure as the density approaches the close packing value, where the system has to be both positionally and orientationally frozen. For $k>1$, we encounter a dramatic change in the equation of state, because the close packing structure becomes unique with perfect nematic ordering along *y* axis. This change manifests in $P/P_\parallel$ vs $\rho$ curves as shown in Fig. 7. Interestingly $P/P_\parallel$ becomes double-peaked for *k*=1.001. The first peak at lower density belongs to mixed ordering as the system can maximize the entropy with more orientational states in mixed order rather than in nematic one. The emergence of the second peak is due to the close packing constraint, which can be realized with perfectly nematic arrangement. The peak of mixed ordering does not move with the density, while the nematic peak moves away from the close packing density ($\rho^*=1$) as *k* increases from 1. At *k*=1.1 the peak of mixed order disappears completely due to the shift of the nematic peak to the direction of lower densities. We can also see that the structural change from mixed to nematic weakens with increasing *k*, which manifests in a decreasing $P/P_\parallel$ peak. These results show that the tendency for mixed ordering weakens suddenly with increasing *k*. The inset of Fig, 7 shows that $\alpha$ is discontinuous with a jump of $\Delta\alpha$=0.4 at *k*=1, i.e. $\alpha$=1.9 (1.5) for *k*>1 (*k*=1). This result shows that even though the angular fluctuations always contribute to the pressure, the contribution of angular fluctuations is higher in the nematic than in the mixed ordering.

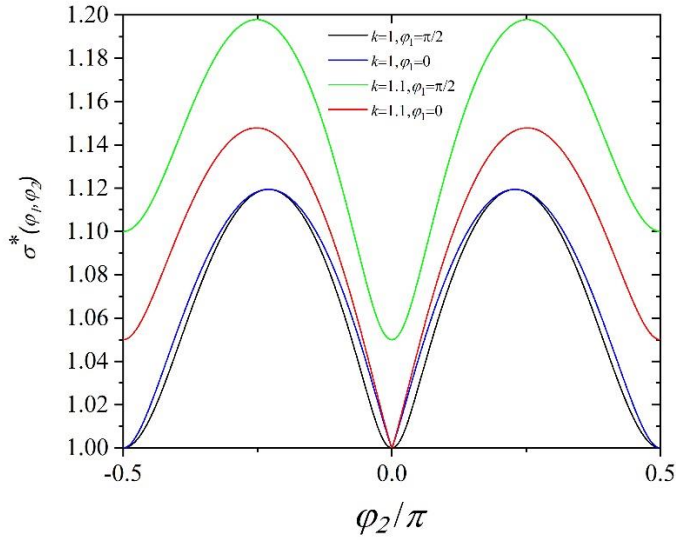

**Figure 6.** Contact distance between two superparticles. The orientation of particle 1 is horizontal ($\varphi_1=\pi/2$) or vertical ($\varphi_1=0$), while the orientation of particle 2 is varying between $-\pi/2$ and $\pi/2$. The results are shown for superdisk and superellipse shapes with $n=2$ and $m=10$. $\sigma^*=\sigma/d$ is the dimensionless contact distance.

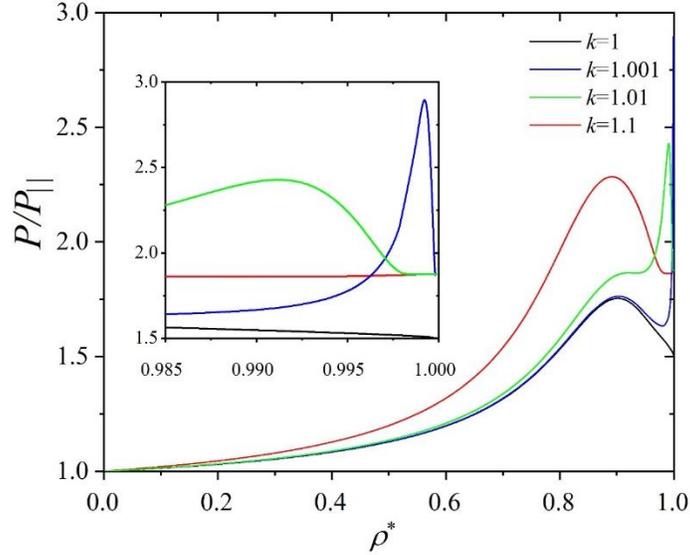

**Figure 7.** The effect of aspect ratio (*k*) on the equation of state of q1D hard superdisks with deformation parameters *n*=2 and *m*=10 in the pressure ratio vs. density plane. The behaviour near close packing is enlarged in the inset. $\rho^*=\rho\, d$ is the dimensionless density.

The fraction of vertical particles behaves differently in mixed and nematic structures as shown in Fig. 8. In the mixed ordering the number of vertical particles is less than the horizontal ones ($x_v<x_h$), while the opposite happens in the nematic structure ($x_v>x_h$). Therefore, $x_v$ is always less than 1/2 in the fluid of superdisks as this fluid forms only mixed structure. The effect of increasing $k$ is that the region of $x_v<1/2$ shrinks, i.e. the mixed ordering is destabilized with respect to nematic ordering. It can be seen in Fig. 8 that the mixed ordering is completely missing for $k=1.1$ as $x_v>1/2$ in the whole density range. This is in accordance with Fig. 7, where only one peak belonging to nematic ordering is present for $k=1.1$ in the $P/P_\parallel$ vs. $\rho$ curve. The ratio of the two peaks of the orientational distribution function behaves very similarly to $x_v$, because $f(0)/f(\pi/2)<1$ in mixed, while $f(0)/f(\pi/2)>1$ in nematic arrangement. Note that the onset of nematic ordering can also be captured with the condition $f(0)/f(\pi/2)=1$. It can be seen in the inset of Fig. 8 that the density corresponding to the condition $f(0)/f(\pi/2)=1$ decreases with increasing $k$, which is in accordance with the behavior of the nematic $P/P_\parallel$ peak.

The tetratic and nematic order parameters are shown for superdisks ($k=1$) and superellipses ($k>1$) in Fig. 9. Both $S_2$ and $S_4$ are higher than zero even for superdisks, which shows that the violation of fourfold rotational symmetry with $n \neq m$ manifests in $S_2>0$. In the dilute regime both order parameters are nearly zero, which means that the structure is almost isotropic (we can call it as quasi-isotropic) as there is enough room for rotation. Therefore the q1D fluid maximizes its entropy with orientational disorder up to intermediate densities. Fig. 9 shows that $S_2$ stays below its possible maximum value ($S_2<1$), while $S_4$ goes to 1 at the close packing density for $k=1$. This highlights the imperfectness of the orientational ordering of superdisks. However, perfect nematic ordering develops for $k>1$, where both $S_2$ and $S_4$ converge to 1 at the close packing density. It can be also seen in the behavior of $S_2$ and $S_4$ that the particles form mixed order for $k\leq1.01$ and the mixed order changes to be nematic at high densites. The position of the mixed-to-nematic structural change occurs, where $S_2$ has local minimum close to the maximal density (see the inset of Fig. 9). This local minimum is completely missing for $k=1.1$, where only nematic ordering develops continuously.

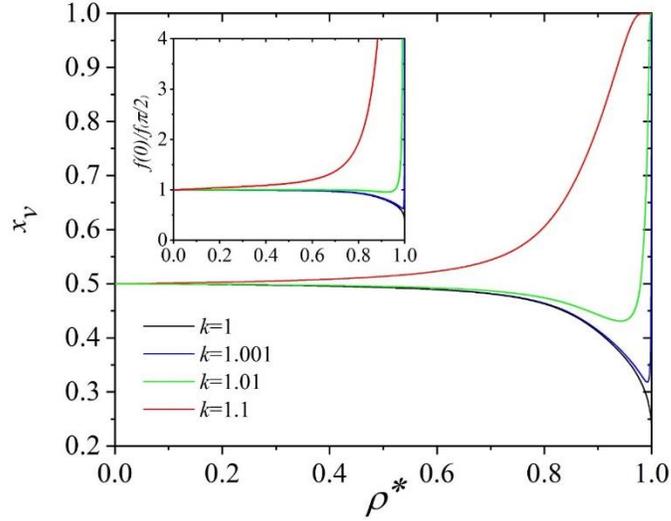

**Figure 8** The effect of aspect ratio ($k$) on the orientational ordering of hard superdisks with $n=2$ and $m=10$. The fraction of vertical particles is shown as a function of dimensionless density ($\rho^*=\rho d$). The inset shows the ratio of orientational distribution functions measured at $\varphi=0$ and $\varphi=\pi/2$. The limiting value of $x_v$ as $\rho^* \to 1$ is about 0.25 for $k=1$, while it is 1 for $k>1$.

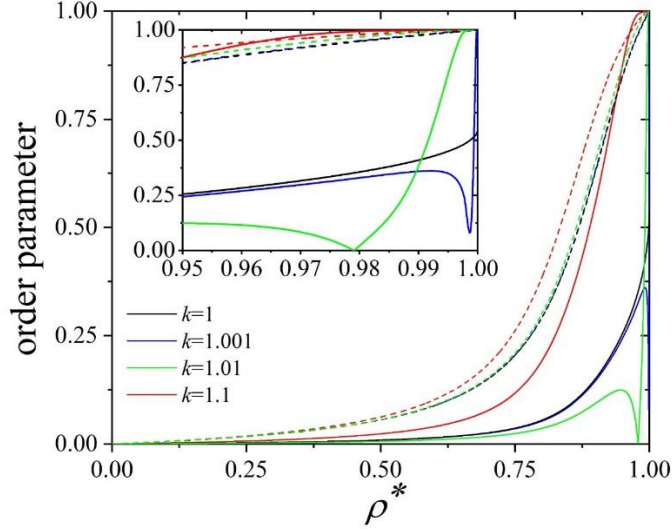

**Figure 9** The effect of aspect ratio on the orientational ordering of hard superdisks with $n=2$ and $m=10$. The nematic (continuous curve) and tetratic (dashed curve) order parameters are shown as a function of dimensionless density ($\rho^*=\rho\, d$). The close packing behaviour is enlarged in the inset.

Angular fluctuations must vanish with approaching the close packing. In the tetratic and mixed structure, the preferred orientations are located at $\varphi_d=0$ and $\pm\pi/2$, while only $\varphi_d=0$ is preferred in nematic (see Fig. 5). To capture all structures, we measure the strength of angular fluctuations with

$$\left\langle \left(\varphi_d - |\varphi|\right)^2 \right\rangle = \int_{-\pi/2}^{\pi/2} d\varphi \, f(\varphi) \left(\varphi_d - |\varphi|\right)^2, \tag{12}$$

where $\varphi_d=\pi/2$ for $\pi/4<|\varphi|\leq\pi/2$ and $\varphi_d=0$ for $0<|\varphi|\leq\pi/4$. As it can be seen in Fig. 10, the signatures of structural changes are present in the behavior of average angular fluctuations, too. For all cases, the average angular fluctuation remains approximately constant up to $P^*\approx 10$, after which it decreases toward zero with a power law. It can be seen that the average angular fluctuation obeys a different power law decay in mixed and nematic structures. The decay is weaker in the mixed state with $P^{-1}$, while it follows a stronger $P^{-1.4}$ dependence in the nematic. For $k=1$ and $k=1.1$ only a single power law dependence is present in accordance with the one-peak structure of $P/P_\parallel$. However, a two-step power-law decay connected with an intermediate exponential change are present for systems with $k=1.001$ and $1.01$. The first power law dependence is related to the mixed ordering, while the second one to the nematic ordering in accordance with the two-peak structure of $P/P_\parallel$. Not surprisingly, the signatures of mixed and nematic ordering are visible in the behaviour of orientational correlation length, too, as shown in Fig. 11. Regarding the structural change from quasi-isotropic to mixed structure, the correlation length starts to grow exponentially in the quasi-isotropic structure, but it saturates at a short length in the mixed ordering for $n=2$ (see left panel of Fig. 11). Interestingly the situation is slightly different for superdisks with $n=4$ (see right panel of Fig. 11), where the mixed structure exhibits a diverging correlation length with pressure according to $\xi \sim P^{0.25}$. This shows that superdisks with $n=2$ are always weakly correlated, while those with $n=4$ become strongly correlated at the vicinity of close packing density. For $k=1.001$ and $1.01$, where quasi-isotropic, mixed and nematic structures are present, the pressure dependence of the orientational correlations in quasi-isotropic and mixed structures are the same as before, but the mixed to nematic structural change is accompanied with exponential grow of $\xi$, changing to a power-law in the nematic as $\xi \sim P^{0.4}$. In the superellipse fluid with $k=1.1$, the exponential dependence changes to be power law as only the quasi-isotropic structure changes to be nematic with increasing pressure.

Finally we show our results for hard superballs having deformations $n=2$ and $m=10$. This system is also degenerate at close packing, because the contact distance is the same for axial-axial, radial-radial and radial-axial configurations. Therefore the particles can orient along the $z$ axis and in the radial $x$-$y$ plane with arbitrary azimuthal angle ($0<\varphi<2\pi$) in the close packing configuration.

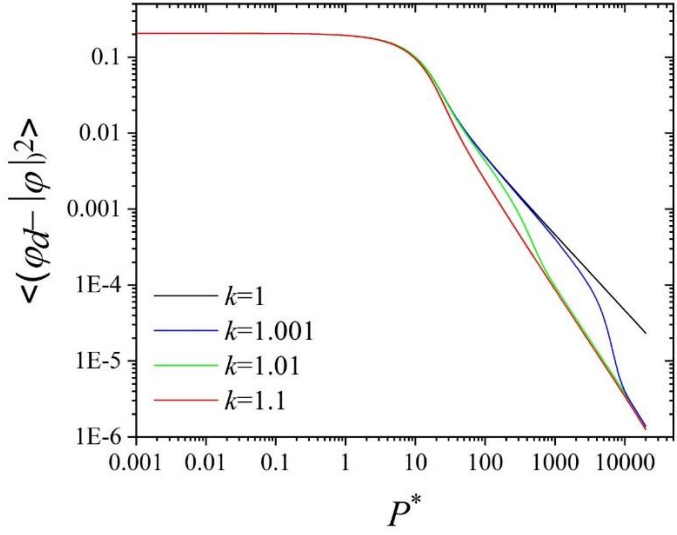

**Figure 10** The effect of aspect ratio on the angular fluctuations. The average of angular fluctuations from the preferred $\varphi_d$ orientation as a function of dimensionless pressure ($P^*=\beta Pd$). The hard superdisk ($k=1$) and superellipse ($k>1$) are characterized with $n=2$ and $m=10$.

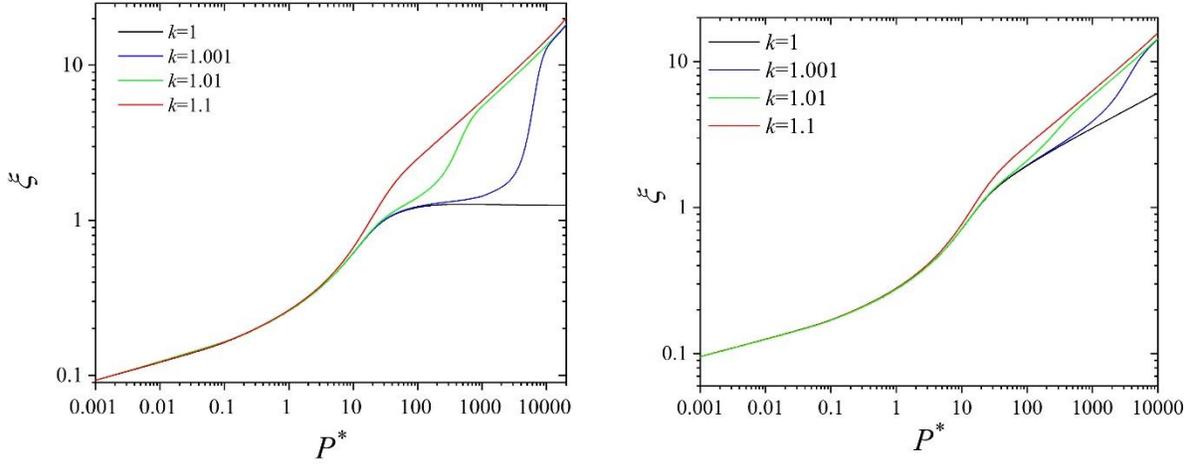

**Figure 11** The effect of aspect ratio on the orientational correlations. The orientational correlation length as a function of dimensionless pressure ($P^*=\beta Pd$). The left panel belongs to $n=2$ and $m=10$, while the right one to $n=4$ and $m=10$.

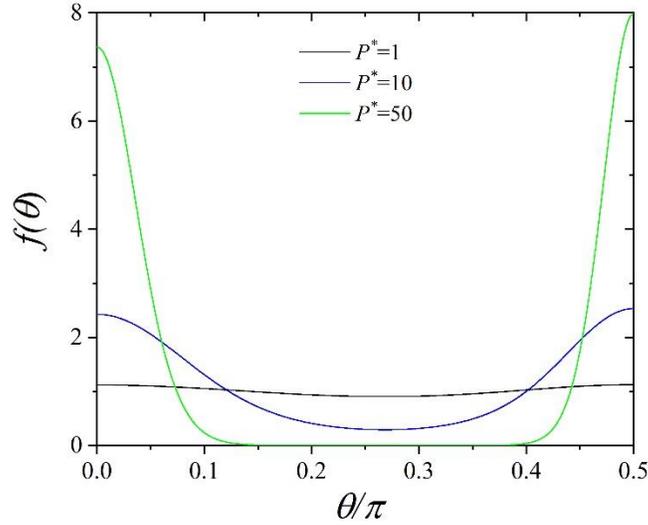

**Figure 12** Ordering of hard superball particles with $n=2$ and $m=10$. Orientational distribution function as a function of polar angle. $P^*=\beta Pd$ is the dimensionless pressure.

To see the similarity between superdisks and superballs, the orientational distribution function of superballs is presented in Fig. 12. We can see that $f$ is neither nematic nor cubatic, because it has a peak at both radial and axial directions with $f(0){\neq}f(\pi/2)$, which is also the feature of mixed ordering of superdisks (see Fig. 5). Therefore, the axial ($\theta=0$) and radial ($\theta=\pi/2$) values of $f$ characterize the structure of superball fluid. The lowest cost in contact distance between two orientationally fluctuating superballs can be realized with radial ordering, which is shown in Fig. 13. It can be seen that about 70% of the superballs are radial in a wide range of densities, which increases further to 100 % at the close packing density. However, this cannot be seen in the ratio of $f(0)$ and $f(\pi/2)$, which is larger than 1 with approaching the close packing density. However, this does not contradict our finding that $x_r>0.5$, because $f(0)$ corresponds only to a single orientation, while $f(\pi/2)$ to all orientations with azimuthal angle $\varphi$ in the interval of $0<\varphi<2\pi$, which is continuum number of angles. Therefore continuum number of orientations wins over the discrete one in the fraction of radial particles, and the two-peak structure of the orientational distribution function does not manifest in the close packing value of $x_r$. The consequence of the 3D rotation of superballs is that the ordering is different from the mixed ordering of 2D superdisks because $x_r$ goes to 1, but $f(0)/f(\pi/2)$ remains finite at the close packing. This behavior does not occur in the nematic state, where $f$ should be single-peaked at $\theta=0$ and $f(0)/f(\pi/2)=\infty$ at the close packing. The behaviour of cubatic ordering is also very different, where there are three mutually perpendicular orientations with the same peak in the orientational distribution function. Therefore, the structure of the superballs can be also considered as mixed ordering, which inherits some of the properties of nematic and cubatic ordering. Finally, we note that pressure ratio ($P/P_\parallel$), the order parameters ($S_2$ and $S_4$) and the average fluctuations in polar angle also support our findings for the mixed ordering of superballs. These and other findings are presented in the supplementary material.

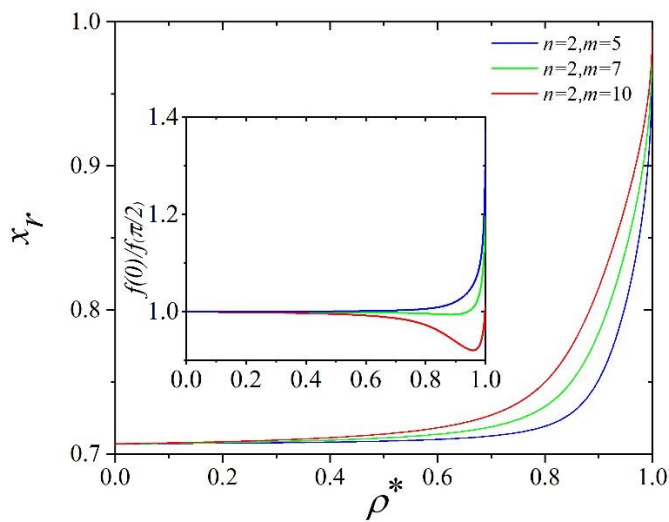

**Figure 13** Fraction of radially arranged superball particles as a function dimensionless density ($\rho^*=\rho\, d$). The inset shows the ratio of orientational distribution function measured at $\theta=0$ and $\theta=\pi/2$.

**Conclusions**

We have presented an exact theoretical study of orientational ordering in quasi-one-dimensional (q1D) hard superparticle fluids, revealing the existence of a novel mixed structure that emerges from subtle geometric asymmetries in the particle shape. The main feature of the mixed structure is that for q1D superdisks it is intermediate between tetratic and nematic structures, while for q1D superballs it resembles to both cubatic and nematic states. To induce mixed order, the shape of the particle must be anisotropic such a way that the close packing configuration remains degenerate. This can be achieved with changing the curvature of neighboring sides (faces) to be different, but the distance between opposite sides (faces) must be the same.

The mixed orientational ordering is characterized by an excess of both parallel and perpendicular orientational ordering in the q1D channel throughout the entire density range. The mixed ordering of superdisks can be distinguished from the other type of orderings with several unique features: (i) the nematic order parameter remains less than one even at close packing, while the tetratic order parameter approaches one, (ii) the fraction of vertical particles is consistently less than 0.5, indicating a preference for horizontal orientations, and (iii) the orientational distribution function exhibits two distinct peaks with unequal heights ($f(0) \neq f(\pi/2)$). The mixed ordering of superballs is slightly different from that of superdisks, because the fraction of radial superballs goes to 1 even if two unequal peaks in the orientational distribution function ($f(0) \neq f(\pi/2)$) survives at the close packing density.

Through systematic variation of two deformation parameters ($n$ and $m$) in the shape of both superdisks and superballs, we have established that the mixed ordering emerges specifically when the neighboring sides (faces) have different curvature ($n \neq m$), while maintaining equal characteristic lengths along the axes of the superparticle ($k=1$). The stability of the mixed structure arises from an entropic gain in the case of superdisks: vertical particles preferring to be the neighbor of horizontal ones due to lower contact distance costs, while horizontal particles show no orientational preference for their neighbors. Therefore the asymmetric excluded distance cost leads to the formation of mixed structure, where the orientational entropy can be higher without giving rise to packing entropy loss. In the case of superballs, even if an asymmetry is present in the excluded distance cost, the orientational entropy selects the radial orientation for the ordering at the close packing. This manifests clearly in the fraction of radial superballs, which converges to 1, while the fraction of axial particles goes to zero in the close packing limit. However, the common feature of mixed ordering for both superdisks and superballs is that $f(0) \neq f(\pi/2)$ and $f(0)/f(\pi/2)$ is finite in the whole density range.

Our stability analysis on q1D superdisk fluid reveals that the mixed ordering is remarkably sensitive to the particle's aspect ratio ($k$). Even minimal deviation from the superdisk shape (e.g. $k=1.001$) initiates a structural change towards nematic ordering. Moreover, the mixed structure completely disappears at $k=1.1$. The aspect ratio sensitivity of the ordering manifests in a double-peaked structure in the pressure ratio ($P/P_\parallel$) versus density curves, where the first peak is due to mixed ordering and the second to nematic one. Even the close packing value of the pressure ratio ($\alpha$) is very sensitive to the value of $k$ as it exhibits a discontinuity at $k=1$. This is due to the fact that the close packing is associated mainly with horizontal particles for $k=1$, while the vertical ones dominate for $k>1$. Therefore, the deformation parameter $n$ plays a dominant role for $k=1$, while it is $m$ for $k>1$. Actually, our results shows that $\alpha=2-1/n$ for $k=1$ and $\alpha=2-1/m$ for $k>1$. These equations explain the jump $\Delta\alpha=0.4$ in $\alpha$ in superdisk fluids with $n=2$ and $m=10$. Note that the observed jump in $\alpha$ also provides clear thermodynamic evidence for the distinct nature of mixed and nematic structures. The average angle fluctuations also exhibit peculiar behavior as they vanish with a power law in $P$ as $P^{1/n-3/2}$ for $k=1$ and $P^{1/m-3/2}$ for $k>1$. Interestingly, a two-step power law decay can be observed for $k=1.001$ and $1.01$, where the mixed order with a decay $P^{1/n-3/2}$ changes to be nematic with a decay $P^{1/m-3/2}$ decay upon compression. However, no sign of mixed order is observed in the behavior of average angular fluctuations as it decays with $P^{1/m-3/2}$ for $k\geq 1.1$. The orientational correlation length further distinguishes the mixed structure from the nematic one. For superdisks with $n=2$ and $m>2$, the correlation length ($\xi$) saturates at finite values, but for superdisks with $n=4$ and $m\geq 4$ it diverges. These behaviors can be unified into a law $\xi \sim P^{1/2-1/n}$, which survive also for $1<k<1.01$. As the structure changes from mixed to nematic, the orientational correlation length diverges as $\xi \sim P^{1/2-1/m}$. For $k=1.1$, the two-step power law divergence is replaced with a single $\xi \sim P^{1/2-1/m}$ dependence. The reason for the replacement of $n$ with $m$ in the above laws can be also understood as the horizontal-horizontal collisions are dominant in mixed order, while the vertical-vertical ones in the nematic order. Interestingly the close packing properties do not depend directly from $k$, but all of them exhibit discontinuous behavior at $k=1$. Regarding the superballs, the mixed ordering can be destabilized with both oblate ($k<1$) and prolate ($k>1$) deformation of the particle. The nematic order along the channel replaces the mixed one for $k<1$, while the planar order with a single peak at $\theta=\pi/2$ becomes stable for $k>1$.

      The discovery of mixed orientational ordering in q1D systems also raises questions about the existence of similar phases in two- and three-dimensional systems under appropriate confinement conditions. The interplay between geometric frustration, entropic effects, and shape anisotropy may yield additional novel phases that have yet to be explored theoretically or experimentally. For example, two-dimensional melting can be more complex with more than two-

steps including new intermediate mesophases and even new solid-solid transitions can arise in systems of hard superdisks having two deformation parameters.

In conclusion, this work demonstrates that q1D confinement can amplify the effects of subtle particle shape asymmetries, leading to the emergence of novel orientational phases. The mixed structure identified here expands our understanding of liquid crystalline ordering and provides new insights into the fundamental role of particle geometry in determining collective behavior of confined colloidal systems.

**Acknowledgement**

S.M. acknowledges financial support from the Alexander von Humboldt Foundation. P.G. and S.V. gratefully acknowledge the financial support of the National Research, Development, and Innovation Office - Grants No. NKFIH K137720, No. 2023-1.2.4-TÉT-2023-00007 and No. TKP2021-NKTA-21.